\documentclass[aps,twocolumn,showpacs,preprintnumbers,amsmath,amssymb]{revtex4}
\usepackage{graphicx}
\usepackage{dcolumn}
\usepackage{bm}
\usepackage{amsmath}
\usepackage{amssymb}
\usepackage{color}
\usepackage{amsmath}
\usepackage{amssymb}
\usepackage{bm}
\usepackage{braket}
\def\dd{\mathrm{d}}
\def\ee{\mathrm{e}}
\def\ii{\mathrm{i}}

\def\ddt{\frac{\partial }{\partial t}}

\def\Hc{\mathrm{H.c.}}
\def\Re{\mathrm{Re}}

\def\epsz{\varepsilon_0}
\def\epsb{\varepsilon_{\mathrm{bg}}}

\def\w{\omega}
\def\wex{\varOmega^{\rm ex}}
\def\ww{\varOmega}
\def\wT{\omega_{\mathrm{T}}}
\def\wLT{\omega_{\mathrm{LT}}}

\def\mex{m_{\mathrm{ex}}}
\def\dampex{{\it \gamma}_{\mathrm{ex}}}
\def\phaseex{{\it \Gamma}_{\mathrm{ex}}}

\def\thick{L}

\def\win{\omega_{\mathrm{in}}}
\def\kin{k_{\mathrm{in}}}

\def\Sic{S_{\mathrm{ic}}}

\def\eV{\mathrm{eV}}
\def\meV{\mathrm{meV}}

\def\cm{\mathrm{cm}}

\def\nm{\mathrm{nm}}
\def\Watt{\mathrm{W}}

\def\dimP{\mathcal{P}}
\def\dimE{\mathcal{E}}
\def\dimF{\mathcal{F}}

\def\G{\mathsf{G}}

\def\oH{\hat{H}}
\def\oHrad{\hat{H}_{\mathrm{rad}}}
\def\oHint{\hat{H}_{\mathrm{int}}}
\def\oHex{\hat{H}_{\mathrm{ex}}}

\def\superphase{\hat{{\mathcal L}}_{\mathrm{phase}}}
\def\superdamp{\hat{{\mathcal L}}_{\mathrm{damp}}}

\def\density{\hat{\rho}}

\def\oex{\hat{b}}
\def\oexd{\hat{b}^{\dagger}}

\def\oph{\hat{a}}
\def\ophd{\hat{a}^{\dagger}}

\def\os{\hat{\sigma}}
\def\osd{\hat{\sigma}^{\dagger}}

\def\oPex{\hat{P}_{\mathrm{ex}}}

\def\oE{\hat{E}}
\def\oEd{\hat{E}^\dagger}
\def\deltaoE{\varDelta\hat{E}}
\def\deltaoEd{\varDelta\hat{E}^\dagger}

\def\oEz{\hat{E}_0}
\def\oEdz{\hat{E}^\dagger_0}
\def\Ein{\mathcal{E}_{\mathrm{in}}}

\def\zz{z_0}
\def\z{z}

\usepackage{prettyref}
\newrefformat{eq}{(\ref{#1})}
\newrefformat{fig}{\ref{#1}}
\def\pref#1{\prettyref{#1}}

\begin{document}
\title{Up-converted photoluminescence induced by radiative coupling between excitons}

\author{Takuya Matsuda}
\email{matsuda@pe.osakafu-u.ac.jp}
\author{Nobuhiko Yokoshi}
\author{Hajime Ishihara}
\email{ishi@pe.osakafu-u.ac.jp}

\affiliation{Department of Physics and Electronics, 
Osaka Prefecture University, Sakai, Osaka 599-8531, Japan}


\begin{abstract}
We propose an unconventional scheme of photoluminescence in a semiconductor thin film, where 
the nonlocal correlation between an excitonic wave and a light wave prominently enhances the exciton--light 
coupling beyond the long-wavelength approximation (the so-called excitonic superradiance regime). 
On the basis of the developed method extending input--output theory, we elucidate atypical photoluminescence effects
due to the strong wave--wave correlation.
In particular, the up-converted photoluminescence based on the coherent superposition of excitons is found to be 
highly efficient, i.e., it can be realized by weak pumping without auxiliary systems 
such as cavities or photonic antennas.
\end{abstract}


\maketitle




Photoluminescence (PL) is one of the fundamental phenomena arising from light--matter coupling, and has been extensively utilized for monitoring the structure of the electronic levels in materials~\cite{pelant12}. PL spectroscopy provides the essential information of matter systems, such as the structure of the levels and oscillator strengths of electronic excited states after reaching quasi-equilibrium. On the other hand, PL is also important for observing the interplay scheme between light and electronic systems. A representative example is PL reflecting the dispersion relation of polaritons and not the thermal equilibrium for bare excitons~\cite{sumi75,ivchenko77,ivchenko89}. In order to access polaritons by photon emission, photons have to be emitted before the exciton--photon coupling is ``dephased", or the excitations relax to quasi-equilibrium. Then, it is necessary to consider polariton accumulation at the bottleneck of the decay~\cite{sumi75}, or confine the electronic system in high-density photons~\cite{imamoglu96,dang98,laussy04,christopoulos07}.

Recently, another scheme that allows PL to access exciton--light coupled modes has been demonstrated in simple semiconductor thin films~\cite{phuong12}. When the film thickness is in the nano-to-bulk crossover regime (a few hundred nanometers), the exciton--light coupling is extremely enhanced by the nonlocal correlation between a light wave and an excitonic center-of-mass (c.m.) wave~\cite{ishi02,ishi04,syouji04,ichimiya09}. Note that the resultant exciton--light coupled state should be distinguished from polaritons because it exhibits an ultrafast radiative decay exceeding the polariton formation, and is in the so-called excitonic superradiance regime~\cite{hanamura88,knoester92,bjork95}. Such a strong coupling beyond the long-wavelength approximation (LWA) leads to unprecedented PL properties. Actually, photon emission simultaneously occurs from more than one exciton--light coupled mode including optically--forbidden ones~\cite{phuong12}.

Here, we focus on a unique feature of the beyond the LWA regime; there exists strong coupling between the different c.m. states of excitons via radiation that causes quantum superposition of these states. The purpose of this paper is to theoretically demonstrate that the control of the superposition provides unconventional and functional aspects of PL properties. In particular, remarkable up-converted PL is revealed, where the up-conversion efficiency is greatly strengthened for appropriate system parameters. Up-converted PL has been an appealing subject not only for its fundamental interest but also for its potential applications such as frequency conversion~\cite{poles99,auzel04,fernandez06,eshlaghi08,paudel11,neupanea13,osaka14}. There are some typical schemes, e.g., those based on phonon-assisted processes~\cite{poles99,auzel04,fernandez06,eshlaghi08}, and on multiphoton processes~\cite{paudel11,neupanea13,osaka14}. Recently, Fern\'ee $et~al.$ reported that the up-conversion range is greatly widened by an unusually--large dephasing rate in a colloidal solution of quantum dots~\cite{fernee07}. The up-conversion in the present work is essentially distinguished from the previous ones in that it is the PL from the exciton--light coupled state with a higher exciton--light coupled mode. The appearance of such a higher state can be explained as follows; a portion of the excited eigenmode is reconstructed to superposed eigenmodes with dephasing into different c.m. exciton states. In addition, the main process can be clearly realized by weak pumping without auxiliary systems such as cavities or photonic antennas.

In the following demonstrations, we treat an explicit geometric model of thin films for numerical calculations in order to discuss the observed results in Ref. ~\onlinecite{phuong12}. The thin film confines the c.m. motion of an exciton perpendicular to the surface and acts as a homogeneous, nondispersive, and nonabsorptive background medium with a dielectric constant $\epsb$ for the excitons. For developing the theoretical method to calculate the luminescence spectra for the nano-to-bulk crossover size regime, it is necessary to take account of the nonlocal optical response of the excitons; thus, we explicitly 
consider the microscopic spatial structures of the excitonic wave and light wave. The Bohr radius of an exciton is assumed to be much smaller than the film thickness $\thick$; therefore, its relative wavefunction is approximated to be the same as that in a bulk system. Thus, neglecting the distortions near the surfaces, we assume the c.m. states of exciton to be simple sinusoidal waves: $\psi_\mu(\z)=\sqrt{2/\thick}\,\sin(k_\mu \z)$,  where $k_\mu=\mu\pi/\thick$ is the quantized wavevector with $\mu=1,2,\cdots$. The transverse energies and the eigenenergies of the exciton are given as $\hbar\wT=3.2022\,\eV$ and $\hbar\wex_\mu=\hbar\wT+(\hbar k_\mu)^2/(2\mex)$, respectively. Here, $\mex$ is the translational mass of the exciton.

The Hamiltonian of the exciton--light coupled system is written as $\oH=\oHex+\oHrad+\oHint$. The Hamiltonian of the excitons is described as $\oHex=\sum_\mu \hbar\wex_\mu \oexd_\mu \oex_\mu$, and $\oex_\mu$ ($\oexd_\mu$) stand for the bosonic annihilation (creation) operator of the $\mu$th exciton state. The Hamiltonian of the radiation field is represented as $\oHrad=\sum_\eta \hbar\ww_\eta \ophd_\eta \oph_\eta$, where $\oph_\eta$ ($\ophd_\eta$) means the annihilation (creation) operator of the $\eta$th photon mode with an energy $\hbar\ww_\eta$. The interaction between the exciton and the radiation field is expressed as $\oHint=-\int\dd\z \oPex(\z)\oE(\z)$. Here, the excitonic polarization operator $\oPex(\z)$ is represented as $\oPex(\z)=\sum_\mu \big(\dimP_\mu(\z)\oex_\mu+\Hc\big)$ by using $\dimP_{\mu}(\z)=\dimP\ \psi_\mu(\z)$ with the dipole moment $\dimP$. The moment $\dimP$ can be estimated by the longitudinal--transverse (LT) splitting energy $\hbar\wLT=\dimP^2/(\epsz\epsb)$. The operator $\oE(\z)=\sum_\eta \alpha_\eta\big(\ii \dimE_\eta(\z)\oph_\eta+\Hc\big)$ denotes the electric field, where $\alpha_\eta=\sqrt{\hbar\ww_\eta/(2\epsz)}$, $\epsz$ is the vacuum permittivity and $\dimE_\eta(\z)$ is the eigenfunction satisfying Maxwell equation without excitonic polarization sources.

As for the excitonic non-radiative processes, we treat them under the Born--Markov approximation. Then, the density matrix of the entire system obeys the following master equation:
\begin{align}\label{eq:master_eq}
\ddt
\density(t)
=
\frac{1}{\ii\hbar}
[\oH,\, \density(t)]
+\superdamp\density(t) 
+\superphase\density(t),
\end{align}
where the non-radiative decay of the exciton states, e.g., due to phonons, is described by $\superdamp\density(t)
=
\sum_{\mu}(\dampex/2)\Big[2\os_{\mu,\mu+1}\density(t)\osd_{\mu,\mu+1}
-\big\{\osd_{\mu,\mu+1} \os_{\mu,\mu+1},\density(t)\big\}\Big]$, and the dephasing is described by 
$\superphase\density(t)
=
\sum_\mu(\phaseex/2)
\Big[\oexd_{\mu}\oex_{\mu},\big[\oexd_{\mu}\oex_{\mu},\density(t)\big]\Big]$. We have introduced the lowering operator for excitons $\os_{\mu,\mu+1}=\oexd_\mu\oex_{\mu+1}$. In the present demonstration, it is assumed that non-radiative transitions only occur between the nearest levels. All non-radiative decay rates are the same $\hbar\dampex=0.01\,\meV$, and the dephasing rates of each exciton state are $\hbar\phaseex=0.2\,\meV$ considering the cryogenic conditions. These assumptions do not affect the essence of the following discussion because the non-radiative decay rates are much smaller than the radiative decay rates under the present conditions, although the quantitative details of the results depend on the values of the non-radiative ones (the influence of $\hbar\dampex$ is examined in Ref.~\onlinecite{SM}). From Eq. \pref{eq:master_eq}, we obtain the simultaneous Heisenberg equations to determine the polarization $\braket{\oex_\mu}$, the excitonic population $\braket{\oexd_\mu\oex_\mu}$, and the correlation $\braket{\oexd_\mu\oex_{\mu'}}$~\cite{cho03}.  The detailed solutions of these equations are given in Ref.~\onlinecite{SM}.

\begin{figure}[h] 
\vspace{-1em}
\includegraphics[width=\linewidth]{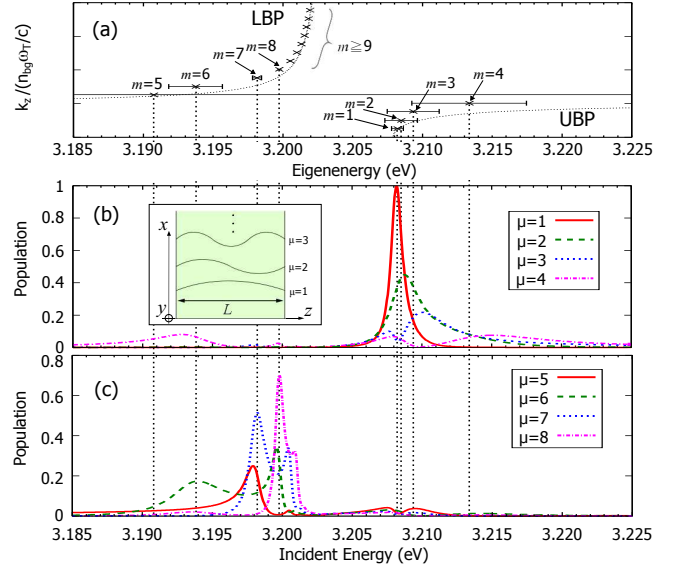}
\caption{\label{fig:fig1}(Color online) (a) The dotted lines represent the dispersion curves of the upper- and lower-branch polaritons (UBP and LBP), and the cross marks indicate the resonance energies of the exciton--light coupled modes (vertical dotted lines) in the film with $\thick=325\,\nm$. The bar length represents the radiative decay rates of the coupled modes. $m$ denotes the index of the original exciton state of each coupled mode. (b,c) The population spectra of different exciton states (in this case $\mu=1,2,\cdots,8$) are plotted
versus the incident energy, and $\mu$ represents the exciton state. Every population spectrum has a complicated structure because of the very large radiative coupling between different excitonic waves. All populations are normalized by the peak value of the population of the exciton states $\mu=1$. The c.m. wavefunctions of the excitons [see inset of (b)] are considered.}
\label{fig1}
\end{figure} 


First, we analyze the excitonic population $\braket{\oexd_\mu\oex_\mu}$. Hereafter, unless otherwise noted, the system is assumed to be a CuCl film with a thickness $\thick=325\,\nm$ modeling the experiment in Ref. \onlinecite{phuong12}, where the dielectric constant is $\epsb=5.59$. As for the incident light, we employ a continuous wave laser whose intensity is $I_\text{in} =c \epsz \Ein^2/2=100\,\Watt/\cm^2$, where $c$ is the speed of light, and $\Ein$ is the amplitude of the incident light. In Fig.~\ref{fig1}(a), we plot the eigenenergies of the exciton--light coupled modes, as denoted by the cross marks, whereas the dotted lines indicate the dispersion relation of the upper- and lower-branch polaritons. It should be noted that the dispersion relation of the exciton--light coupled modes ($m=1,2,\dots$) deviates from that of the polaritons in the bulk system owing to the finite radiative width~\cite{ishi04}. The radiative width becomes enlarged near the LT splittings, and it reaches $96\,\meV$ for the coupled mode $m=5$. Figures~\ref{fig1}(b,c) show the populations of the different excitonic states $\braket{\oexd_\mu\oex_\mu}$ plotted versus the incident light energy. One can see that each population has peaks reflecting the eigenenergies of the exciton--light coupled modes (vertical dotted lines) due to the radiation-mediated coupling between the different c.m. states of the exciton.
This coupling is described with the terms appearing in the Heisenberg equations (see, \cite{SM}):
\begin{align}
Z_{\mu,\mu'}
=
-\int\dd\z\int\dd\z'\
\dimP_\mu^*(\z)\G(\z,\z')\dimP_{\mu'}(\z'),
\label{eq:radiative_coupling}
\end{align}
where $\G(\z,\z')$ is the Green's function for Maxwell equation in the vacuum/film/vacuum structure~\cite{chew95}. This term represents the coupling between the exciton states mediated by the radiation fields.
Because of this interaction, each population $\braket{\oexd_\mu\oex_\mu}$ exhibits structures due to the resonances at
the exciton--light coupled modes or the interference between them.
Notice that the peaks slightly deviate from the eigenenergies especially above the LT splitting, because the adjacent modes overlap owing to the radiative width. By focusing on the c.m. states $\mu=4, 5$, we can clearly see that their spectra remain visible on both the sides of the LT splitting. This means that the wave--wave coupling between the excitons and the radiation field causes the coherent superposition of different exciton c.m. states over the LT splitting.


Subsequently, we introduce a calculation method to investigate the PL properties. Here we extend the conventional input--output theory to include the nonlocal correlation inside the film. By using the Green's function $\G(\zz,\z')$ for the radiation field, the electric field can be written as \cite{wubs04}
\begin{align}
\oE(\zz,t)
&=
\oEz(\zz,t)
+\sum_\mu
\int\dd\z'\
\G(\zz,\z')\dimP_\mu(\z')\oex_\mu(t),
\label{output}
\end{align} 
where $\zz$ is an arbitrary position. Here the incident field $\oEz(\zz,t)$ corresponds to the one in the absence of the resonant contributions of the excitons. Then, the input pump field is the plane wave with the amplitude $\Ein$, which is connected to satisfy the Maxwell boundary condition in the vacuum/film/vacuum structure. By using the output radiation field in Eq.~(\ref{output}), we calculate the first-order correlation function $\braket{\oEd(\zz,t)\oE(\zz,t+\tau)}$ according to the quantum regression theorem~\cite{walls94,carmichael99}, and decompose it into a coherent component and an incoherent one. The coherent component corresponds to elastic scattering, whereas the incoherent component provides inelastic scattering, namely ``the luminescence". Then, the incoherent component is written as $\braket{\deltaoEd(\zz,t)\deltaoE(\zz,t+\tau)}$=$\braket{\oEd(\zz,t)\oE(\zz,t+\tau)}-\braket{\oEd(\zz,t)}\braket{\oE(\zz,t+\tau)}$, where the last term on the right-hand side corresponds to the coherent component. Under steady-state conditions, the PL spectrum on the transmission side is obtained by calculating the Fourier transform of the incoherent correlation function: 
\begin{align}\label{eq:spe}
\Sic^\text{T}(\w)
&=
\Re
\Big[
\int_0^\infty\dd\tau
\frac{\ee^{\ii\w\tau}}{\pi}\braket{\deltaoEd_\text{T}(\zz,0)\deltaoE_\text{T}(\zz,\tau)}
\Big], 
\end{align} 
where the field $\oE_\text{T}(\zz,t)$ shows the electric field on the transmission side. Hereafter we normalize the PL intensity by $\wLT/\Ein^2$.

\begin{figure}[t] 
\vspace{-1em}
\includegraphics[width=\linewidth]{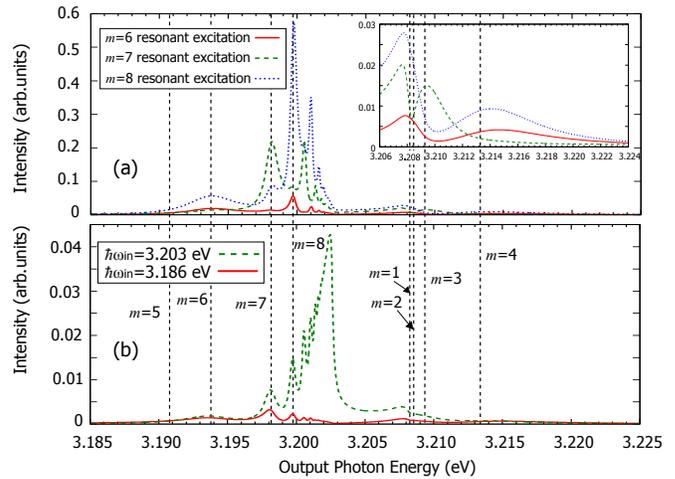}
\caption{\label{fig:fig2} (a) The PL intensity is plotted versus the output photon energy. The incident energy $\hbar\win$ is tuned to the resonance energies of the coupled modes $m=6$, $m=7$, and $m=8$ in Fig.1 (a). (b) The PL intensity is plotted under the condition in which the incident energy $\hbar\win$ is tuned to $3.186\,\eV$ (experimental condition in Ref.~\onlinecite{phuong12}) and $3.203\,\eV$ (stopband region). The vertical dotted lines represent the eigenenergies of the exciton--light coupled modes.}
\label{fig2}
\end{figure} 


In Fig.~\ref{fig2}(a), we plot the PL intensities versus the output energy for different incident light energies which are set to be resonant with the exciton--light coupled modes below the LT splitting, i.e., $m=\{6,7,8\}$ in Fig.~\ref{fig1}(a). One can see that luminescence beyond the LT splitting appears, i.e., the light energies are up-converted. Here, we remark that the energy conservation law should hold true for the whole system in steady state, even though the luminescence includes up-converted components. Indeed, the present model and calculation scheme guarantee this law, which is verified in Ref.~\onlinecite{SM}.

The essential factor of the observable PL up-conversion
is the exciton--light coupling, extremely enhanced by the
nonlocal correlation beyond the LWA. The mechanism of PL up-conversion is interpreted to be a result of quantum superposition between a resonantly-excited exciton--light coupled mode and the one above the LT splitting. To be precise, a small portion of the excited coupled mode is reconstructed into multiple superpositions of the coupled modes by the enhanced exciton--light coupling under slow dephasing to the exciton c.m. states ($\mu=1,2,\dots$). Actually, the PL appears to reflect the eigenenergies of the coupled modes (vertical dotted lines), unlike conventional PL~\cite{pelant12}. This means that the PL signal can plausibly identify the exciton--light interactions.

Further, photon emission is affected by the parity-selection rule of the exciton c.m. state. For example, if the coupled mode $m$=7 is resonantly excited, a signal predominantly appears from the odd-numbered states. Conversely, if the coupled mode $m=\{6,8\}$ is excited, the even-numbered states provide the primary contribution to PL. This is consistent with our criterion that the proposed up-converted PL originates from neither phonon assisted, multiphoton processes nor huge dephasing but from the radiation-mediated superposition of the exciton--light coupled modes. In addition, there must exist a finite probability amplitude, even inside the LT splitting, due to the overlap between the adjacent coupled modes to superpose the states on both sides of the LT splitting. Indeed, the coherent superposition also affects the PL properties inside the LT splitting. In Fig~\ref{fig2}(b), the PL spectrum for $\hbar \win=3.203\,\eV$ (green dashed line) that lies inside the LT splitting. One can apparently see that PL signals are found inside the LT splitting, which never occurs in a bulk sample.

In order to compare the calculation with the existing experiment using a $389$-$\nm$ coherent light source (Ti:sapphire)~\cite{phuong12}, we investigate the case where the incident light energy is set to be $\hbar \win=3.186\,\eV$ [see the red solid line in Fig.~\ref{fig2}(b)]. Although the incident energy is far detuned from the coupled modes, we confirm that the signal is small but appears at the eigenenergies above the LT splitting. 
It should be remarked that we can actually find a similar signal, namely, up-converted PL, in the experiment described in Ref.~\onlinecite{phuong12}~\cite{signal}. Although the verification of the origin of this signal by theory considering the detailed experimental conditions will be a subject of future study, this fact convinces us of the reality of the present proposal. As for the up-conversion efficiency, we evaluate the photon-number efficiency as
\begin{align}\label{eq:eff}
\eta_{\hbar\w_\text{a}}^{\hbar\w_\text{b}}
\equiv
\frac{1}{\Ein^2}
\int_{\w_\text{a}}^{\w_\text{b}}\dd\w\ 
\Sic^\text{T}(\w).
\end{align}
Here, we count the photons emitted from the coupled state $m=4$ above the LT splitting, i.e., between $\hbar \w_\text{b}=3.220\,\eV$ and  $\hbar \w_\text{a}=3.212\,\eV$. Although the incident intensity is as small as $100\,\Watt/\cm^2$ and off-resonant, the evaluated photon-number efficiency is $\eta_{3.212\,\eV}^{3.220\,\eV}=0.077\,\%$, which is sufficiently large to be detected in an experiment.

\begin{figure}[tb] 
\vspace{-1em}
\includegraphics[width=\linewidth]{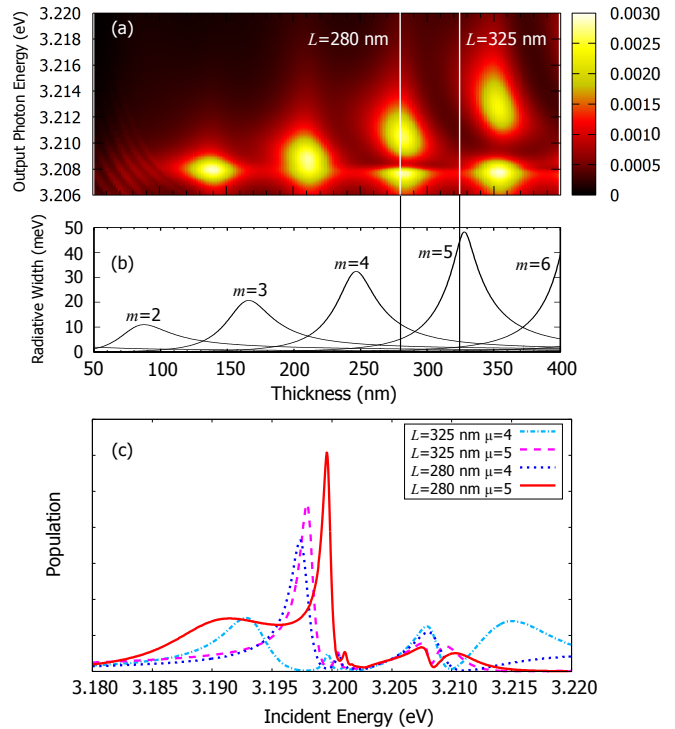}
\caption{\label{fig:fig3} (a) The PL intensity is plotted as functions of the film thickness and output photon energy. The incident energy is tuned to $3.186\,\eV$. (b) The film thickness dependence of the calculated radiative width. (c) The population spectra of different exciton states (in this case $
\mu=4,~5$ at thicknesses of $\thick=280\,\nm,~325\,\nm$) is plotted
versus the incident energy}
\label{fig3}
\end{figure} 


Because the strong nonlocal correlation in the film requires synchronization between the excitonic wave and the light wave, the up-conversion efficiency should depend on the film thickness $\thick$. In Fig.~\ref{fig3}(a), we show the PL intensity plotted versus the thickness and output energy when $389$-$\nm$ coherent laser light ($\hbar \win =3.186\,\eV$) is injected. As is expected, the intensity is found to exhibit a strong thickness dependence. One may consider that the larger radiative width is favorable for efficient up-converted PL. However, when comparing Figs.~\ref{fig3}(a,b), the peaks in the PL intensity do not directly coincide with those of the radiative width. We therefore investigate the excitation populations of the c.m. states of the exciton for the two thicknesses (see Fig.~\ref{fig3}(c)). One is the thickness used in Figs.~\ref{fig1} and~\ref{fig2} ($\thick =325\,\nm$), whereas the other is the thickness that provides a strong PL intensity ($\thick =280\,\nm$). By calculation, the photon-number efficiency for the thickness $\thick=280\,\nm$ is $\eta_{3.209\,\eV}^{3.217\,\eV}=0.207\,\%$, which is more than twice as large as the one for $\thick=325\,\nm$. By considering the reason for the enhancement, we now focus on the region around $\hbar \win =3.19\,\eV$, where the eigenenergy of the coupled mode $m=5$ lies. One can see that the population of the exciton c.m. state $\mu=5$ has a broad peak for the thickness $\thick=280\,\nm$ (red solid line), whereas the very large radiative width saturates the peak for $\thick=325\,\nm$ (pink dashed line). This leads to a larger population for the coupled mode $m=5$ at $\hbar \win =3.186\,\eV$ for $\thick=280\,\nm$, and an enhancement in up-converted PL. This means that efficient up-converted PL by off-resonant pumping can be obtained 
in the experiment by invoking the theoretical design of the system considering the populations of the exciton c.m. states.


In summary, we have examined photoluminescence (PL) in a simple semiconductor thin film and proposed an unconventional type of up-converted PL in the excitonic superradiance regime. In order to treat the nonlocal correlation beyond the long-wavelength approximation (LWA), we extend conventional input--output theory~\cite{walls94,carmichael99}.  In our numerical calculations of the PL spectra, we have found that a portion of the incident energy is up-converted, and the efficiency is enhanced by choosing the sample size and pumping frequency. The favorable conditions for enhancement clearly reflect the parity combination of the c.m. wavefunctions. The up-converted PL originates from the coherent superposition of different exciton--light coupled modes in the resonantly populated excitons. The strong exciton--light coupling via the nonlocal correlation beyond the LWA greatly enhances the superposed component energetically far from the excitation energy. This is why the main process can be clearly realized by weak pumping without auxiliary systems such as cavities or photonic antennas.

It should be noted that the thin film geometry used for the present demonstrations that exhibits a suggestive signal in the experiment \cite{phuong12} is a very familiar system, and further, the proposed mechanism has generality. However, the excitonic superradiance can be enhanced in planar microcavities~\cite{odani93,savona95}. Thus, it might be possible to design more efficient systems if one examines such systems over a widely expanded volume~\cite{deych07,averkiev09}. By controlling the coherent light--matter coupling in designed structures, the proposed up-converted PL will lead to future studies of functional PL for novel light sources as well as sensitive probes of light--matter coupling.

We thank M. Ashida and M. Ichimiya for valuable discussions, and T. M. thanks H. Ajiki for fruitful discussion. This work was partially supported by a Grant-in-Aid for Scientific Research (A) No. 24244048 from Japan and by the Japan Society for Promotion of Science (JSPS).

\appendix

\begin{widetext}
\section{DETAILED SOLUTIONS OF HEISENBERG EQUATIONS}
In this section, we present detailed solutions of the Heisenberg equations to obtain the expectation values of the polarization $\braket{\oex_{\mu}}$, the population $\braket{\oexd_\mu\oex_{\mu}}$ and the correlation $\braket{\oexd_\mu\oex_{\mu'}}$ under steady-state conditions. In the main text, we consider the Hamiltonian as
\begin{align}
\oH
=
\oHex+\oHrad+\oHint.
\end{align}
The Hamiltonian $\oHex$ describes the excitonic system as 
\begin{align}
\oHex
=
\sum_\mu\hbar\wex_\mu\
\oexd_\mu\oex_\mu,
\end{align}
where $\oex_\mu$ $(\oexd_\mu)$ stand for the bosonic annihilation (creation) operator of an exciton state $\mu$ and $\wex_\mu$ is its eigenfrequency. The Hamiltonian $\oHrad$ represents the radiation field as 
\begin{align}
\oHrad
=
\sum_\eta \hbar\ww_\eta\
\ophd_\eta\oph_\eta,
\end{align}
where $\oph_\eta$ $(\ophd_\eta)$ stand for the annihilation (creation) operator of the $\eta$th photon state with an energy $\hbar\ww_\eta$. The coupling Hamiltonian $\oHint$ signifies the interaction between the exciton and the radiation field as
\begin{align}
\oHint
=
-\int\dd\z\ 
\oPex(\z)\oE(\z).
\end{align}
The excitonic polarization operator $\oPex(\z)$ is represented as 
\begin{align}
\oPex(\z)
=
\sum_\mu(\dimP_\mu(\z)\oex_\mu+\Hc)
\end{align}
by using the coefficient $\dimP_\mu(\z)=\dimP~\psi_\mu(\z)$ with the dipole moment $\dimP$. The electric field operator $\oE(\z)$ is denoted as
\begin{align}\label{eq:demba}
\oE(\z)
=
\sum_\eta\alpha_\eta(\ii\dimE_\eta(\z)\oph_\eta+\Hc),
\end{align}
where $\alpha_\eta = \sqrt{\hbar\ww_\eta/(2\epsz)}$ and $\dimE_\eta(\z)$ is the eigenfunction satisfying the Maxwell equation without excitonic polarization sources. By solving the Heisenberg equation of motion for the exciton and the radiation field, we can derive the electric field by using Green's function for the radiation field~\cite{wubs04} given by
\begin{align}\label{eq:radiation}
\oE(\z,t)
&=
\oEz(\z,t)
+\sum_\mu
\int\dd\z'\
\G(\z,\z')\dimP_\mu(\z')\oex_\mu(t).
\end{align}

We consider the density matrix of the whole system obeying the master equation in the main text given by
\begin{align}\label{eq:master}
\ddt
\density(t)
& =
\frac{1}{\ii\hbar}
[\oH , \density(t)]
+\superdamp\density(t)+\superphase\density(t),
\end{align}
where the non-radiative decay of the exciton states is described as 
\begin{align}
\superdamp\density(t)
& =
\sum_{\mu}
\frac{\dampex}{2}
\Big[
2\os_{\mu,\mu+1}\density(t)\osd_{\mu,\mu+1}
-\bigl\{
\osd_{\mu,\mu+1}\os_{\mu,\mu+1}, \density(t)
\bigr\}
\Big],
\end{align}
and the dephasing is given by
\begin{align}
\superphase\density(t)
& =
\sum_{\mu}
\frac{\phaseex}{2}
\Big[
\oexd_{\mu}\oex_{\mu}\
\big[
\oexd_{\mu}\oex_{\mu}, \density(t)
\big]
\Big].
\end{align}
We introduce the lowering operator of the excitons $\os_{\mu,\mu+1}=\hat{b}_{\mu}^{\dagger}\hat{b}_{\mu+1}$. We assume that the non-radiative transitions only occur between nearest levels. All non-radiative decay rates are the same $\dampex$, and all dephasing rates are the same $\phaseex$. From the master equation in Eq.~(\ref{eq:master}), we can obtain the following equations:
\begin{align}
\ddt
\braket{\oex_{\mu}(t)}
&=
-(\ii\wex_{\mu}+\dampex/2+\phaseex/2)
\braket{\oex_{\mu}(t)}
+\frac{\ii}{\hbar}\int\dd\z \
\dimP^*_{\mu}(\z)\braket{\oE(\z,t)},
\label{eq:pol}
\end{align}
\begin{align}
\ddt
\braket{\oexd_{\mu}(t)\oex_{\mu}(t)}
&=
-\dampex\braket{\oexd_{\mu}(t)\oex_{\mu}(t)}
+\dampex\braket{\oexd_{\mu+1}(t)\oex_{\mu+1}(t)}
\nonumber\\
&\quad
+\frac{\ii}{\hbar}\int\dd\z \
\big(
\braket{\oexd_{\mu}(t)}\dimP^*_{\mu}(\z)\braket{\oEz(\z,t)}
-\dimP_{\mu}(\z)\braket{\oEdz(\z,t)}\braket{\oex_{\mu}(t)}
\big)
\nonumber\\
&\quad
-\frac{\ii}{\hbar}
\sum_\nu
\braket{\oexd_{\mu}(t)\oex_\nu(t)}Z_{\nu,\mu}
+\frac{\ii}{\hbar}\sum_{\lambda}Z^*_{\mu,\lambda}\braket{\oexd_{\lambda}(t)\oex_{\mu}(t)},
\label{eq:popu}
\end{align}
\begin{align}
\ddt
\braket{\oexd_{\mu}(t)\oex_{\mu'}(t)}
&=
\Big[
{\ii}(\wex_{\mu}-\wex_{\mu'})-\dampex-\phaseex
\Big]
\braket{\oexd_{\mu}(t)\oex_{\mu'}(t)}
\nonumber\\
&\quad
+\frac{\ii}{\hbar}\int\dd\z \
\big(
\braket{\oexd_{\mu}(t)}\dimP^*_{\mu'}(\z)\braket{\oEz(\z,t)}
-\dimP_{\mu}(\z)\braket{\oEdz(\z,t)}\braket{\oex_{\mu'}(t)}
\big)
\nonumber\\
&\quad
-\frac{\ii}{\hbar}
\sum_{\nu}
\braket{\oexd_{\mu}(t)\oex_{\nu}(t)}Z_{\nu,\mu'}
+\frac{\ii}{\hbar}
\sum_{\lambda}\ 
Z^*_{\mu,\lambda}
\braket{\oexd_{\lambda}(t)\oex_{\mu'}(t)},
\label{eq:cor}
\end{align}
where 
\begin{align}
Z_{\mu,\mu'}
=
-\int\dd\z\int\dd\z'\ \dimP^*_{\mu}(\z)\G(\z,\z')\dimP_{\mu'}(\z').
\end{align}
This term represents the radiation-mediated coupling between excitons, which plays a central role in our work. This radiative coupling between excitons causes the coherent superposition of different exciton states mediated by the strong exciton--photon coupling via nonlocal correlation. 
By considering the self-consistent coupling between these equations and the electric field~\cite{cho03}, the expectation values can be obtained. By substituting Eq.~(\ref{eq:radiation}), we can rewrite Eq.~(\ref{eq:pol}) under the rotating wave approximation as
\begin{align}\label{eq:self_consistent_eq}
\sum_{\mu'}
M_{\mu,\mu'}(\w)
\braket{\oex_{\mu'}(\w)}
&=
\int\dd\z \
\dimP^*_{\mu}(\z)\braket{\oEz(\z,\w)}
\end{align}
where 
\begin{align}
M_{\mu,\mu'}(\w)
\equiv
\hbar(\wex_{\mu}-\w-\ii\dampex/2-\ii\phaseex/2)\delta_{\mu,\mu'}+Z_{\mu,\mu'}. 
\end{align}
This simultaneous equation set in Eq.~(\ref{eq:self_consistent_eq}) is solved by the inverse matrix $\bm{\mathsf{N}}(\w)=[\bm{\mathsf{M}}(\w)]^{-1}$ as
\begin{align}\label{eq:b}
\braket{\oex_{\mu}(\w)}
&=
\sum_{\mu'}
N_{\mu,\mu'}(\w)
\int\dd\z \
\dimP^*_{\mu'}(\z)\braket{\oEz(\z,\w)}.
\end{align}
The incident light is assumed to be a monochromatic continuous wave laser as
\begin{align}
\braket{\oEz(\z,\w)}
=
E_0(\z)\ \delta(\w-\win).
\end{align}
Then, we rewrite Eq.~(\ref{eq:b}) as
\begin{subequations}
\begin{align} 
\braket{\oex_{\mu}(\w)}
&=
\beta_{\mu}(\w) \ \delta(\w-\win),
\\
\beta_{\mu}(\w)
&\equiv
\sum_{\mu'}
N_{\mu,\mu'}(\w)
\int\dd\z \
\dimP^*_{\mu}(\z)E_0(\z).
\end{align}
\end{subequations}
By performing Fourier transform, we can obtain an analytical expression of $\braket{\oex_{\mu}(t)}$ under steady-state conditions as
\begin{align}
\braket{\oex_{\mu}(t)}
=
\beta_{\mu}(\win)\ \ee^{-\ii\win t}.
\end{align}
By using this result, the simultaneous equations Eq.~(\ref{eq:popu}) and Eq.~(\ref{eq:cor}) can be analytically solved, and the population $\braket{\oexd_{\mu}(t)\oex_{\mu}(t)}$ and the correlation $\braket{\oexd_{\mu}(t)\oex_{\mu'}(t)}$ are obtained under steady-state conditions.

\section{ENERGY CONSERVATION LAW}

In this section, we verify the energy conservation law, i.e., show that the incident light energy matches the total energy in the system. The total energy consists of the elastically and inelastically scattered photon energies, non-radiative decay energy, and dephasing energy. To calculate each type of energy, we follow the technique based on the input--output theory~\cite{koshino11}. First, we define the spectrum of the incident monotonic continuous wave laser $E_\text{in}(\z,t)=\dimE_\text{in}\ \ee^{\ii(\kin\z-\win t)}$ as
\begin{align}
S_\text{in}(\w)
=
\Re
\Big[
\frac{1}{\pi}
\int_{0}^{\infty}\dd\tau\
\ee^{\ii\w\tau}
\{\dimE_\text{in}\ \ee^{\ii\kin\z}\}^*\dimE_\text{in}\ \ee^{\ii(\kin\z-\win\tau)}
\Big]
=
\Ein^2\ \delta(\w-\win).
\end{align}
The spectrum of coherent components of the output radiation field on the transmission side is expressed as
\begin{align}\label{eq:STc}
S^\text{T}_\text{c}(\w)
&=
\Re
\Big[
\int_0^{\infty}\dd\tau\
\frac{\ee^{\ii\w\tau}}{\pi}
\braket{\oEd_\text{T}(\zz,0)}\braket{\oE_\text{T}(\zz,\tau)}
\Big],
\end{align}
and that of one of the incoherent components, which corresponds to the photoluminescence (PL) spectrum (Eq.(4) in the main text), is expressed as
\begin{align}\label{eq:STic}
S^\text{T}_\text{ic}(\w)
&=
\Re
\Big[
\int_0^{\infty}\dd\tau\
\frac{\ee^{\ii\w\tau}}{\pi}
\braket{\deltaoEd_\text{T}(\zz,0)\deltaoE_\text{T}(\zz,\tau)}
\Big].
\end{align}

Subsequently, we calculate the spectra of non-radiative decay and dephasing. In this work, we have employed the Markov approximation for the operators $\os_{\mu,\mu+1}=\hat{b}_{\mu}^{\dagger}\hat{b}_{\mu+1}$ and $\hat{B}_\mu=\oexd_\mu\oex_\mu$. Then, on the basis of the input--output theory, we can represent the output field as  
\begin{align}
\hat{D}^{\mu-1,\mu}_\text{damp}(t)
&=
-\ii\sqrt{\dampex}\
\os_{\mu-1,\mu}(t),
\\
\hat{D}^\mu_\text{phase}(t)
&=
-\ii\sqrt{\phaseex}\
\hat{B}_\mu(t).
\end{align}
In addition, within the Markov approximation, the number flux of the photons can be written as
\begin{align}
\dimF(z,t)
& =
\frac{1}{\hbar\win}
Q(z,t),
\end{align} 
where the power of the light $Q(z,t)$ in quantum optics is~\cite{wubs04}
\begin{align}
Q(z,t)
&=
2\epsz c
\braket	{\oEd(z,t)\oE(z,t)}.
\end{align}
Therefore, under steady-state conditions, the spectra of the non-radiative decay and dephasing are, respectively, written as
\begin{subequations}
\begin{align}
S_\text{damp}(\w)
&=\frac{1}{\it \Lambda}
\Re
\Big[
\sum_\mu
\int_0^{\infty}\dd\tau\
\frac{\ee^{\ii\w\tau}}{\pi}
\braket{\{\hat{D}^{\mu-1,\mu}_\text{damp}(0)\}^\dagger\hat{D}^{\mu-1,\mu}_\text{damp}(\tau)}
\Big]
\nonumber\\
&=
\frac{\dampex}{\it \Lambda}~
\Re
\Big[\sum_\mu
\int_0^{\infty}\dd\tau\
\frac{\ee^{\ii\w\tau}}{\pi}
\braket{\os_{\mu,\mu-1}(0)\os_{\mu-1,\mu}(\tau)}
\Big],
\\
S_\text{phase}(\w)
&=\frac{1}{\it \Lambda}
\Re
\Big[\sum_\mu
\int_0^{\infty}\dd\tau\
\frac{\ee^{\ii\w\tau}}{\pi}
\braket{\{\hat{D}^\mu_\text{phase}(0)\}^\dagger\hat{D}^\mu_\text{phase}(\tau)}
\Big]
\nonumber\\
&=
\frac{\phaseex}{\it \Lambda}~
\Re
\Big[\sum_\mu
\int_0^{\infty}\dd\tau\
\frac{\ee^{\ii\w\tau}}{\pi}
\braket{\hat{B}^\dagger_\mu(0)\hat{B}_\mu(\tau)}
\Big],
\end{align}
\end{subequations}
where  ${\it \Lambda}=2\epsz c/(\hbar\win)$.

In order to verify the energy conservation law, we have to establish the correctness of the following equation:
\begin{align}
\int_{-\infty}^{\infty}\dd\w\
\hbar\w\ S_\text{c}(\w)
+\int_{-\infty}^{\infty}\dd\w\
\hbar\w\ S_\text{ic}(\w)
+\int_{-\infty}^{\infty}\dd\w\
\hbar\w\ S_\text{damp}(\w)
+\int_{-\infty}^{\infty}\dd\w\
\hbar\w\ S_\text{phase}(\w)
&=
\hbar\w_{\rm in} \Ein^2.
\end{align}
In Table~\ref{tb:ECL}, we show each component in percentage with respect to $\hbar\win \Ein^2$, i.e., $P_{i}=\int_{-\infty}^{\infty}\dd\w\ \hbar\w\ S_{i}(\w)/(\hbar\win \Ein^2)\times 100$. We see that the dephasing part $P_\text{phase}$ sometimes becomes negative. Such a negative $P_\text{phase}$ appears when the incident light energy is lower than the transverse exciton energy and the system can continuously absorb energy from the environment. This is discussed in detail in Ref. \onlinecite{koshino11}.  One can see that there is tiny mismatch between $P_{\rm total}$ and $100 \%$. In order to check the validity of our numerical calculations, we examine the following quantity
\begin{eqnarray}
\delta
= \eta_{\hbar\omega_\text{a}}^{\hbar\omega_\text{b}}
\frac{\Ein^2 (\hbar \omega_{\rm out} -\hbar \omega_{\rm in})}
{\hbar\w_{\rm in} \Ein^2}\times 100~~ (\%).
\end{eqnarray}
The quantity $\eta$ is the PL up-conversion efficiency from $\hbar \w_{\rm in}$ to the peak energy $\hbar \w_{\rm out}$, where the output PL spectrum spreads between $\hbar\w_\text{a}$ and $\hbar\w_\text{b}$. The quantity $\delta$ roughly estimates the effect of the up-conversion on the total energy. For example, in case of the film thickness $\thick=325~\nm$ and input energy $\hbar\win=3.186~\eV$, the up-conversion efficiency evaluated in the main text is $\eta_{3.212~\eV}^{3.220~\eV}=0.077~\%$, i.e., $\delta =0.0725~\%$. This is much larger than the maximum mismatch ($\sim 0.0004~\%$). Therefore, we can confirm that the mismatch is caused by numerical error and the energy conservation law holds true even though a portion of the incident light is up-converted.

\begin{table}[t]
\caption{Energy components in percentage for various film thicknesses and incident energies.}
\label{tb:ECL}
\begin{tabular}{ccccccc}
\hline\hline
$\thick~(\nm)$ & $\hbar\win~(\eV)$ & $P_\text{c}~(\%)$ & $P_\text{ic}~(\%)$ & $P_\text{damp}~(\%)$
& $P_\text{phase}~(\%)$ & $P_\text{total}~(\%)$
\\
\hline
$280~(\nm)$ & $3.1860~(\eV)$ & $97.4253~(\%)$ & $2.4362~(\%)$ & $0.1499~(\%) $
& $-0.0114~(\%)$ & $100.0000~(\%)$
\\
$325~(\nm)$ & $3.1860~(\eV)$ & $98.5195~(\%)$ & $1.4048~(\%)$ & $0.0808~(\%)$ 
& $-0.0052~(\%)$ & $99.9999~(\%)$
\\
$325~(\nm)$ & $3.1936~(\eV)$ & $88.7898~(\%)$ & $10.5592~(\%)$ & $0.6716~(\%)$ 
& $-0.0207~(\%)$ & $99.9999~(\%)$
\\
$325~(\nm)$ & $3.1981~(\eV)$ & $69.3738~(\%)$ & $28.6470~(\%)$ & $1.9680~(\%)$
& $0.0108~(\%)$ & $99.9996~(\%)$
\\
$325~(\nm)$ & $3.1997~(\eV)$ & $58.2861~(\%)$ & $38.4866~(\%)$ & $3.2309~(\%)$ 
& $-0.0036~(\%)$ & $100.0000~(\%)$
\\
$325~(\nm)$ & $3.2030~(\eV)$ & $94.8785~(\%)$ & $4.7041~(\%)$ & $0.4134~(\%)$ 
& $0.0039~(\%)$ & $99.9999~(\%)$
\\
\hline\hline
\end{tabular}
\end{table}

\section{EFFECT OF THE NON-RADIATIVE DECAY BETWEEN EXCITON STATES}

\begin{figure}[bth]
\includegraphics[width=12cm]{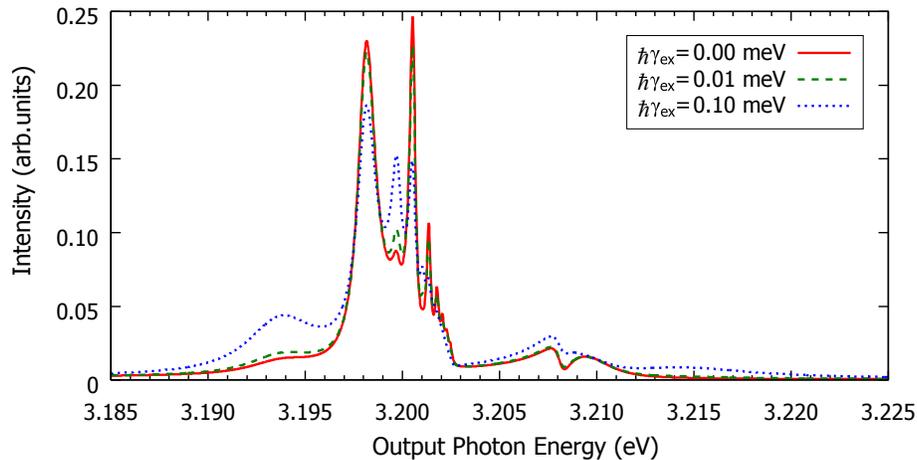}
\caption{The PL intensity is plotted as a function of the output photon energy. The incident energy $\hbar\win$ in is tuned to the resonance energy of the exciton--light coupled mode $m = 7~(3.1981~\eV)$  in case of the film thickness $\thick=325~\nm$. The red solid line represents the spectrum in case of zero non-radiative decay rate $\hbar\dampex=0~\meV$. The green dashed line represents the spectrum in case of the non-radiative decay rate $\hbar\dampex=0.01~\meV$ in the main text in Fig.~2(a). The blue dotted line represents the spectrum in case of a larger non-radiative decay rate $\hbar\dampex=0.1~\meV$. The numerical results indicate that finite non-radiative decay does not change the essential profile of the proposed up-converted PL effect.} 
\label{fig:figS1}
\end{figure}

In this section, we examine the effect of the non-radiative decay between exciton states on PL spectra. We perform calculations where non-radiative decay rate is zero and is comparable to dephasing rate $\hbar\phaseex=0.2~\meV$. For instance, Fig.~\ref{fig:figS1} shows the PL spectra in case of the input energy $\hbar\win=3.1981~\eV$ and film thickness $\thick=325~\nm$. The red solid line represents the spectrum for $\hbar\dampex=0~\meV$, the green dashed line represents the spectrum for $\hbar\dampex=0.01~\meV$ (Fig.~2(a) in the main text), and the blue dotted line represents the spectrum for $\hbar\dampex=0.1~\meV$. In case of a larger non-radiative decay, the spectral shape of the green dashed line in Fig.~\ref{fig:figS1} is slightly different compared with those of the other lines. However, the luminescent peaks appear at almost the same energies, even beyond the LT splitting. Therefore, we confirm that the mechanism of PL up-conversion does not change with non-radiative decay between exciton states.

\end{widetext}


\begin{thebibliography}{10}

\bibitem{pelant12} I. Pelant and J. Valenta, {\it Luminescence Spectroscopy of
Semiconductors}, (Oxford University Press Inc., New York, 2012).

\bibitem{sumi75} H. Sumi, Solid State Commun. {\bf 17}, 701-704 (1975).

\bibitem{ivchenko77} E. L. Ivchenko, G. E. Pikus, B. S. Razbirin, and A. I. Starukhin, JETP {\bf 45}, 1172 (1977).

\bibitem{ivchenko89} E. Ivchenko, A. Selkin, A. Abdukadyrov, M. Sazhin, and N. Yuldashev, Opt. Spectrosc. {\bf 67}, 496 (1989).

\bibitem{imamoglu96} A. Imamo\ifmmode \breve{g}\else \u{g}\fi{}lu, R. J. Ram, S. Pau, and Y. Yamamoto, Phys. Rev. A {\bf 53}, 4250 (1996).
 
\bibitem{dang98} L. S. Dang, D. Heger, R. Andr\'e, F. B\oe{}uf, and R. Romestain, Phys. Rev. Lett. {\bf 81}, 3920 (1998).
 
\bibitem{laussy04} F. P. Laussy, G. Malpuech, A. Kavokin, and P. Bigenwald, Phys. Rev. Lett. {\bf 93}, 016402 (2004).
 
\bibitem{christopoulos07} S. Christopoulos, G. Baldassarri H\"oger von H\"ogersthal,  A. J. D. Grundy, P. G. Lagoudakis, A. V. Kavokin, J. J. Baumberg, G. Christmann, R. Butt\'e, E. Feltin, J. -F. Carlin, and N. Grandjean, Phys. Rev. Lett. {\bf 98}, 126405 (2007).

\bibitem{phuong12} L. Q. Phuong, M. Ichimiya, H. Ishihara, and M. Ashida, Phys. Rev. B {\bf 86}, 235449 (2012).

\bibitem{ishi02} H. Ishihara, K. Cho, K. Akiyama, N. Tomita, Y. Nomura, and T. Isu, Phys. Rev. Lett. {\bf 89}, 017402 (2002).

\bibitem{ishi04} H. Ishihara, J. Kishimoto, and K. Sugihara, J. Lumin. {\bf 108}, 343 (2004).
 
\bibitem{syouji04} A. Syouji, B. P. Zhang, Y. Segawa, J. Kishimoto, H. Ishihara, and K. Cho, Phys. Rev. Lett. {\bf 92}, 257401 (2004).
 
 \bibitem{ichimiya09} M. Ichimiya, M. Ashida, H. Yasuda, H. Ishihara, and T. Itoh, Phys. Rev. Lett. {\bf 103}, 257401 (2009).
 
\bibitem{hanamura88} E. Hanamura, Phys. Rev. B {\bf 38}, 1228 (1988).

\bibitem{knoester92} J. Knoester, Phys. Rev. Lett. {\bf 68}, 654 (1992).
  
\bibitem{bjork95} G. Bj\"ork, S. Pau, J. M. Jacobson, H. Cao, and Y. Yamamoto, Phys. Rev. B {\bf 52}, 17310 (1995).

\bibitem{poles99} E. Poles, D C. Selmarten, O I. Mi\'ci\'c, and A J. Nozik, Appl. Phys. Lett. {\bf 75}, 971 (1999).

\bibitem{auzel04} F. Auzel, Chem. Rev. (Washington, D.C.) {\bf 104}, 139 (2004).

\bibitem{fernandez06} J. Fernandez, A. J. Garcia-Adeva, and R. Balda, Phys. Rev. Lett. {\bf 97}, 033001 (2006).

\bibitem{eshlaghi08} S. Eshlaghi, W. Worthoff, A. D. Wieck, and D. Suter, Phys. Rev. B {\bf 77}, 245317 (2008).

\bibitem{paudel11} H P. Paudel, L. Zhong, K. Bayat, M F. B, S. Smith, C. Lin, C. Jiang, M T. Berry, and P. Stanley May, J. Phys. Chem. C {\bf 115}, 19028 (2011).

\bibitem{neupanea13} B. Neupanea, L. Zhao, and G. Wang, Nano Lett. {\bf 13}, 4087 (2013).

\bibitem{osaka14} Y. Osaka, N. Yokoshi, M. Nakatani, and H. Ishihara,
Phys. Rev. Lett. {\bf 112}, 133601 (2014).

\bibitem{fernee07} M. J. Fern\'ee, P. Jensen, and H. Rubinsztein-Dunlop, Appl. Phys. Lett. {\bf 91}, 043112 (2007).


\bibitem{cho03} K. Cho, {\it Optical Response of Nanostructures: Microscopic Nonlocal Theory}, (Springer Series in Solid-State Sciences, 2003) pp 6-23.

\bibitem{SM} See Supplemental Material.

\bibitem{chew95} W. C. Chew, {\it Waves and Fields in Inhomogeneous Media},
(IEEE, New York, 1995) .

\bibitem{wubs04} M. Wubs, L. G. Suttorp, and A. Lagendijk, Phys. Rev.
A {\bf 70}, 053823 (2004).

\bibitem{walls94} D. F. Walls and G. J. Milburn, {\it Quantum Optics}, (Springer, New York, 1995).

\bibitem{carmichael99} H. J. Carmichael, {\it Statistical Methods in Quantum Optics 1: Master Equations and Fokker-Planck Equation}, (Springer, Berlin, 1998).

\bibitem{koshino11} K. Koshino, Phys. Rev. A {\bf 84}, 033824 (2011).

\bibitem{signal} The relevant signal can be found in Fig. \ref{fig2}(b) in Ref. \onlinecite{phuong12}
for 3.186 eV-ps excitation. The corresponding peak structure obviously reflects 
the characteristic interference among exciton--light coupled states. 

\bibitem{odani93} K. Odani, Y. Ohfuti, and K. Cho, Solid State Commn. {\bf 87}, 507 (1993).

\bibitem{savona95} V. Savona, L. C. Andreani, P. Schwendimann and A. Quantropani, Solid State Commn. {\bf 93}, 733 (1995).

\bibitem{deych07} L. I. Deych, M. V. Erementchouk, A. A. Lisyansky, E. L. Ivchenko, and M. M. Voronov, Phys. Rev. B {\bf 76}, 075350 (2007).

\bibitem{averkiev09} N. S. Averkiev, M. M. Glazov, and A. N. Poddubnyi, JETP {\bf 108}, 836 (2009).


\end{thebibliography}
\end{document}